\documentclass[%
 superscriptaddress,
 amsmath,amssymb,
 aps,
onecolumn,
preprint,
titlepage,
longbibliography
]{revtex4-1}

\usepackage{graphicx}
\usepackage{dcolumn}
\usepackage{bm}
\usepackage{natbib}
\usepackage{color}

\begin{document}


\title{Higher-order realizable algebraic Reynolds stress modeling\\based on the square root tensor}

\author{Kazuhiro Inagaki}
 \email{kinagaki@iis.u-tokyo.ac.jp}
\affiliation{%
Institute of Industrial Science, The University of Tokyo, Tokyo, Japan
}

 \author{Taketo Ariki}%
\affiliation{%
Department of Aerospace Engineering, Tohoku University, Miyagi, Japan
}%

%
\author{Fujihiro Hamba}%
\affiliation{%
Institute of Industrial Science, The University of Tokyo, Tokyo, Japan
}%


\date{\today}




\allowdisplaybreaks[1]

\begin{abstract}
In this study, realizable algebraic Reynolds stress modeling based on the square root tensor [Phys. Rev. E \textbf{92}, 053010 (2015)] is further developed for extending its applicability to more complex flows. In conventional methods, it was difficult to construct an algebraic Reynolds stress model satisfying the realizability conditions when the model involves higher-order nonlinear terms on the mean velocity gradient. Such higher-order nonlinear terms are required to predict turbulent flows with three-dimensional mean velocity. The present modeling based on the square root tensor enables us to make the model always satisfy the realizability conditions, even when it involves higher-order nonlinearity. To construct a realizable algebraic Reynolds stress model applicable to turbulent flows with three-dimensional mean velocity, a quartic-nonlinear eddy-viscosity model is proposed. The performance of the model is numerically verified in a turbulent channel flow, a homogeneous turbulent shear flow, and an axially rotating turbulent pipe flow. The present model gives a good result in each turbulent flow. Note that the mean swirl flow in an axially rotating turbulent pipe flow is reproduced because the present model involves cubic nonlinearity. Such a higher-order realizable algebraic Reynolds stress model, involving quartic nonlinearity on the mean velocity, is expected to be useful in numerically stable predictions of turbulent flows with three-dimensional mean velocity.

\end{abstract}

\pacs{Valid PACS appear here}
\maketitle


\section{\label{sec:level1}Introduction}

Turbulent flows are ubiquitous in real-world fluid flows. Despite remarkable advancements in modern computational technology, most of real-world turbulent flows of high-Reynolds number are still out of reach of direct numerical simulation (DNS). In this sense, the Reynolds averaged Navier-Stokes (RANS) modeling, which treats the statistically averaged quantities of turbulent flows, is a useful tool in predicting high-Reynolds-number turbulent flows.

In a variety of RANS modeling, Reynolds stress transport modeling (RSM, also referred to as second-order closure modeling) is expected to have a potential to give a good prediction for several complex turbulent flows. This is because RSM explicitly calculates the production, dissipation, and diffusion of the Reynolds stress. However, RSM has a problem of realizability; namely, it sometimes gives unphysical predictions such as negative normal stresses \cite{hanjaliclaunderbook,lumley1978}. Schumann \cite{schumann1977} showed three inequality conditions, referred to as the realizability conditions for the Reynolds stress $R_{ij} = \langle u_i' u_j' \rangle$, where $u_i'$ denotes the velocity fluctuation. They are written as follows:
\begin{subequations}
\begin{gather}
R_{\alpha \alpha} \ge 0,
\label{eq:1a} \\
R_{\alpha \alpha} R_{\beta \beta} \ge R_{\alpha \beta}^2,
\label{eq:1b} \\
\mathrm{det} \left[ R_{ij} \right] \ge 0,
\label{eq:1c}
\end{gather}
\end{subequations}
where $\alpha,\beta = 1,2,3$, $\mathrm{det} [ R_{ij} ] = \epsilon_{ij\ell} \epsilon_{abc} R_{ia} R_{jb} R_{\ell c}/6$, and $\epsilon_{ij\ell}$ is the alternating tensor. Note that summations are not taken for Greek indices. The above conditions, Eqs.~(\ref{eq:1a})--(\ref{eq:1c}), are mathematically rigorous and provide a physical importance. For example, the first condition, Eq. (1a), prevents an excessively anisotropic condition caused by the negative normal stresses $R_{\alpha \alpha} < 0$ and the second condition, Eq. (1b), gives the upper bound of the Reynolds shear stress. These lead to the numerical stability of the model. Hence, the realizability conditions given by Eqs.~(\ref{eq:1a})--(\ref{eq:1c}) form a fundamental guideline for the development of physically consistent and numerically stable RANS modeling. However, many simple RANS models, even for the standard eddy-viscosity model, do not necessarily satisfy the realizability conditions. 

For RSM, realizability is often discussed in terms of the time development of the Lumley's flatness $F \{= \mathrm{det} [R_{ij}/(R_{\ell\ell}/3)]\}$ around $F=0$. Here, $F=0$ indicates that turbulence field involves only two-component velocity fluctuation, which is referred to as two-component limit (TCL) condition. In this context, two types of realizability conditions are argued \cite{popebook}; one is the strong realizability, 
\begin{align}
\left. \frac{\mathrm{d} F}{\mathrm{d} t} \right|_{F=0} = 0, \ \ 
\left. \frac{\mathrm{d}^2 F}{\mathrm{d} t^2} \right|_{F=0} > 0,
\label{eq:2}
\end{align}
and the other is the weak realizability,
\begin{align}
\left. \frac{\mathrm{d} F}{\mathrm{d} t} \right|_{F=0} > 0.
\label{eq:3}
\end{align}
Durbin and Speziale \cite{ds1994} gave a stochastic analysis on the weak realizability of the conventional RSM in terms of Langevin equations. Craft and Launder \cite{cl1996,cl2001} proposed highly sophisticated RSM by considering TCL condition in the vicinity of the solid wall. On further development of the realizability conditions for RSM, Girimaji \cite{girimaji2004} discussed and proposed a new constraint on the pressure--strain correlation, and its utility was demonstrated by Mishra and Girimaji \cite{mg2014}.

Although RSM has an advantage that it involves production, dissipation, and diffusion process of the Reynolds stress, its numerical cost is still large because RSM, by nature, demands to solve six individual transport equations for the Reynolds stress. To reduce the numerical cost while retaining the advantages of RSM, the algebraic Reynolds stress modeling (ARSM, or the so-called eddy-viscosity type modeling) is a useful method. ARSM gives an algebraic expression for the Reynolds stress instead of numerically integrating its transport equation. In case of explicit algebraic Reynolds stress modeling (EARSM) which is constructed based on RSM, the realizability of the based RSM is tightly connected to the property of the proposed model. However, when we give an algebraic expression independent of RSM, we should concern Eqs.~(\ref{eq:1a})--(\ref{eq:1c}) instead of Eqs.~(\ref{eq:2}) or (\ref{eq:3}). In this paper, we discuss the realizability conditions for ARSM independent of a specific RSM. Note that the third condition (\ref{eq:1c}) is equivalent to the positive semidefiniteness of the Lumley's flatness, $F \ge 0$. Therefore, ARSM satisfying Eqs.~(\ref{eq:1a})--(\ref{eq:1c}) ensures the weak realizability at least.

To date, many studies on ARSM have been proposed (see, e.g., Ref. \cite{pope1975,tsdia,taulbee1992,gs1993,szl1993,szl1995,girimaji1996,akn1997,wj2000,hamba2001,ajl2003,ariki2019}). However, only a few studies on ARSM consider the realizability conditions. A representative study of realizable ARSM was conducted by Shih \textit{et al}. \cite{szl1993,szl1995}. They constructed ARSM incorporated with quadratic nonlinearity on the mean velocity gradient. Note that the model proposed by Shih \textit{et al}. \cite{szl1993,szl1995} is a quadratic-nonlinear model, and hence, is adequate for turbulent flows with two-dimensional mean velocity \cite{pope1975}. Such flows denote the flows in which the mean velocity field is settled in two dimensions, such as turbulent flows in a plane channel or a straight pipe. Hereafter, we simply refer to turbulent flows with two- or three-dimensional mean velocity as two- or three-dimensional flows, respectively. Some other ARSM which consider the realizability \cite{akn1997,hamba2001,ajl2003} are also quadratic-nonlinear models. However, it was pointed out that cubic nonlinearity is essential for predicting the mean swirl or tangential velocity in an axially rotating turbulent pipe flow, which is an example of the three-dimensional flow \cite{syb2000}. The realizable ARSM involving cubic or higher-order nonlinearity is required to make numerically stable predictions of three-dimensional flows of high-Reynolds number.

Generally, the conventional realizable ARSM can hardly be extended to the models incorporated with higher-order nonlinearity on the mean velocity gradient. Recently, Ariki \cite{ariki2015sqrt} proposed an alternative approach to
realizable ARSM by introducing the square root tensor of the Reynolds stress. In other words, all of eigenvalues of the Reynolds stress constructed by this modeling must be positive semidefinite. Although this approach is a sufficient condition that the model is realizable, it provides a rigorous realizability in a systematic procedure. In the present study, we further develop the square root modeling method, aiming at its application to more complex turbulent flows. Considering the application to three-dimensional flows, a model involving quartic nonlinearity on the mean velocity gradient is proposed and its performance is numerically verified in a turbulent channel flow, a homogeneous turbulent shear flow as basic two-dimensional flow, and an axially rotating turbulent pipe flow as an example of the three-dimensional flow.

The rest of this paper is organized as follows. In Sec.~\ref{sec:level2}, we review the conventional realizable turbulence modeling. Here, we also give an introduction and further development of the turbulence modeling based on the square root tensor \cite{ariki2015sqrt}. In Sec.~\ref{sec:level3}, we construct a quartic-nonlinear model of the Reynolds stress and examine the basic property of the model for the unidirectional shear flow. In Sec.~\ref{sec:level4}, numerical verifications of the proposed model are given. A summary is given and conclusions are discussed in Sec~\ref{sec:level5}.

\section{\label{sec:level2}Modeling the Reynolds stress based on the square root tensor}

\subsection{\label{sec:level2a}Realizability and conventional turbulence modeling}

The simplest and frequently used model for the Reynolds stress is the linear eddy-viscosity model:
\begin{align}
R_{ij} = \frac{2}{3} K \delta_{ij} - \nu_\mathrm{T} S_{ij}.
\label{eq:4}
\end{align}
where $K (= \langle u_i' u_i' \rangle/2))$ denotes the turbulent energy, $\nu_\mathrm{T}$ denotes the eddy viscosity, and $S_{ij}$ denotes the mean strain tensor, which is written in an inertial frame of the Cartesian coordinates as
\begin{align}
S_{ij} = \frac{\partial U_i}{\partial x_j} + \frac{\partial U_j}{\partial x_i}.
\label{eq:5}
\end{align}
Here, $U_i$ denotes the mean velocity. In case of the standard $K$-$\varepsilon$ model, $\nu_\mathrm{T}$ is written as
\begin{align}
\nu_\mathrm{T} = C_\nu \frac{K^2}{\varepsilon}
\label{eq:6}
\end{align}
where $\varepsilon = \nu \langle (\partial u_i' /\partial x_j)^2 \rangle$ denotes the dissipation rate of the turbulent energy and $C_\nu$ is the model constant often optimized as $C_\nu = 0.09$ in a turbulent channel flow \cite{yoshizawabook,popebook}. In this model, $R_{xx}$ is written as
\begin{align}
R_{xx} = \frac{2}{3} K - \nu_\mathrm{T} \frac{\partial U_x}{\partial x} 
= \frac{2}{3} K \left( 1- \frac{3}{2} C_\nu \frac{K}{\varepsilon} \frac{\partial U_x}{\partial x} \right).
\label{eq:7}
\end{align}
As seen from Eq.~(\ref{eq:7}), $R_{xx}$ can take a negative value when $(3C_\nu/2) (K/\varepsilon) \partial U_x/\partial x > 1$, which violates Eq.~(\ref{eq:1a}). Such an unphysical behavior of the Reynolds stress can cause numerical instability. An easy approach to rectify this problem is to extend the constant $C_\nu$ to a functional of the mean strain rate, e.g., $C_\nu \to C_\nu/(1 + C_\nu' K^2 S^2/\varepsilon^2)$, where $S^2 = S_{ij} S_{ij}$. Such a functionalization of the coefficient is a primitive method for the realizable modeling. It is also known that the eddy-viscosity model with $C_\nu = 0.09$ often overestimates the turbulent energy production rate in homogeneous turbulent shear flows \cite{yoshizawabook}. In this sense, the eddy-viscosity coefficient $\nu_\mathrm{T}$ should be generally modeled as a functional form in terms of $S^2$ or other scalar variables in addition to $K$ and $\varepsilon$. In this course, a systematic way of functionalization of the eddy-viscosity coefficient is a matter of modeling.

For the generalization of ARSM, the Reynolds stress is often expanded by the mean velocity gradient \cite{pope1975}, which leads to the following nonlinear eddy-viscosity model:
\begin{align}
R_{ij} & = \frac{2}{3} K \delta_{ij} - \nu_\mathrm{T} S_{ij}
\nonumber \\
& \hspace{1em}
+ \zeta_{SS} \left( S_{ia} S_{aj} - \frac{1}{3} S^2 \delta_{ij} \right)
+ \zeta_{SW} \left( S_{ia} W_{aj} + S_{ja} W_{ai} \right)
+ \zeta_{WW} \left( W_{ia} W_{aj} + \frac{1}{3} W^2 \delta_{ij} \right)
\nonumber \\
& \hspace{1em}
+ \zeta_{SSW} \left( S_{ia} S_{ab} W_{bj} + S_{ja} S_{ab} W_{bi} \right)
+ \zeta_{SWW} \left( S_{ia} W_{ab} W_{bj} + S_{ja} W_{ab} W_{bi} - \frac{2}{3} S_{ab} W_{bc} W_{ca} \delta_{ij} \right)
\nonumber \\
& \hspace{1em} + \cdots,
\label{eq:8}
\end{align}
where $\zeta$'s are dimensional coefficients, $W^2 = W_{ij} W_{ij}$, and $W_{ij}$ denotes the mean absolute vorticity tensor, which is written in an inertial frame of the Cartesian coordinates as
\begin{align}
W_{ij} = \frac{\partial U_i}{\partial x_j} - \frac{\partial U_j}{\partial x_i}.
\label{eq:9}
\end{align}
For a representative of realizable ARSM, Shih \textit{et al}. \cite{szl1993,szl1995} constructed a simple expression for the Reynolds stress in which $\nu_\mathrm{T}$ and $\zeta_{SW}$ are retained but other $\zeta$'s are neglected. In their modeling, the functional form of $\nu_\mathrm{T}$ and $\zeta_{SW}$ were determined to satisfy the realizability conditions for simple shear or rotating turbulent flows. Hamba \cite{hamba2001} discussed a realizable ARSM through the statistical theory for inhomogeneous turbulence referred to as the two-scale direct-interaction approximation (TSDIA) \cite{tsdia}. Abe \textit{et al}. \cite{akn1997} modified the ARSM proposed by Gatski and Speziale \cite{gs1993} to satisfy the realizability conditions for simple flows and Abe \textit{et al}. \cite{ajl2003} further developed its model with careful consideration on the near wall TCL condition. Note that all of them are quadratic-nonlinear eddy-viscosity models. Although these models can capture the anisotropic property of the Reynolds stress in turbulent shear flows owing to the quadratic-nonlinear terms, it is almost impossible to extend these models to higher-order nonlinear ones because the freedom for the dimensional coefficients $\zeta$'s is too large to analytically determine their form. However, Speziale \textit{et al}. \cite{syb2000} pointed out that cubic nonlinearity is essential for predicting the mean swirl or tangential velocity in an axially rotating turbulent pipe flow, which is an example of the three-dimensional flow. Furthermore, the Reynolds stress should be affected not only by the mean velocity gradient but also the turbulent helicity \cite{yy1993,yb2016,inagakietal2017} or the time history effect of the mean strain \cite{speziale1987,hd2008,ariki2019}. The conventional realizable modeling method is hardly extended to such a more general turbulent flow.

\subsection{\label{sec:level2b}Modeling based on the square root tensor}

Ariki \cite{ariki2015sqrt} presented another possibility of realizable ARSM by introducing the square root tensor of the Reynolds stress $\sqrt{R}_{ij}$, which is defined as
\begin{align}
R_{ij} = \sqrt{R}_{ia} \sqrt{R}_{ja}.
\label{eq:10}
\end{align}
Note that $\sqrt{R}_{ij}$ does not signify the square root of each component of the Reynolds stress $R_{ij}$, i.e., $\sqrt{R}_{ij} \neq \sqrt{R_{ij}}$. A physical interpretation of the square root tensor is discussed in Appendix \ref{sec:a}. In this modeling, we choose the square root tensor $\sqrt{R}_{ij}$ as a target of modeling instead of $R_{ij}$ itself. It should be emphasized that the ARSM proposed by this methodology always satisfies the realizability conditions given by Eqs.~(\ref{eq:1a})--(\ref{eq:1c}). In other words, all of eigenvalues of the Reynolds stress are always positive semidefinite. A primitive model for $\sqrt{R}_{ij}$ is an eddy-viscosity type model:
\begin{align}
\sqrt{R}_{ij} = \gamma_0 \delta_{ij} - \gamma_S S_{ij}.
\label{eq:11}
\end{align}
The model given by Eqs.~(\ref{eq:10}) and (\ref{eq:11}) yields the following model of the Reynolds stress:
\begin{align}
R_{ij} = \gamma_0^2 \delta_{ij} - 2 \gamma_0 \gamma_S S_{ij} + \gamma_S^2 S_{ia} S_{ja}.
\label{eq:12}
\end{align}
Thus, even a linear model for $\sqrt{R}_{ij}$ corresponds to a quadratic-nonlinear expression for $R_{ij}$ on the mean velocity gradient The linear eddy-viscosity model can be derived when we add $-\gamma_{SS} S_{ia}S_{aj}$ to Eq.~(\ref{eq:11}) and choose $\gamma_{SS}$ to vanish $S_{ia} S_{aj}$ term in $R_{ij}$. Let us write a general expression for the model of $\sqrt{R}_{ij}$ as 
\begin{align}
\sqrt{R}_{ij} = \gamma_0 \delta_{ij} + \sum_\mathrm{m} \gamma_\mathrm{m} T^\mathrm{m}_{ij},
\label{eq:13}
\end{align}
where $\{ T_{ij}^\mathrm{m} \}$ indicate arbitrary algebraic model terms for $\sqrt{R}_{ij}$, such as $S_{ij}$, $W_{ij}$, or higher-order tensors composed by them. In a previous study by Ariki \cite{ariki2015sqrt}, $\{ T_{ij}^\mathrm{m} \}$ were restricted to symmetric real tensors, but this was not a necessary condition. In this study, $\{ T_{ij}^\mathrm{m} \}$ are allowed to include antisymmetric real tensors.

\subsection{\label{sec:level2c}Further development of square root modeling}

As a further development of square root modeling \cite{ariki2015sqrt}, we propose a systematical approach to determine the functional form of $\gamma_\mathrm{m}$'s following Hamba \cite{hamba2001}. As the trace of the Reynolds stress must correspond to the turbulent energy $K = \langle u_i' u_i' \rangle/2$ as $R_{ii} = 2K$, the model given by Eqs.~(\ref{eq:10}) and (\ref{eq:13}) should satisfy
\begin{align}
2K = R_{ii} = \sqrt{R}_{ij} \sqrt{R}_{ij} = 
3 \gamma_0^2 + 2 \sum_\mathrm{m} \gamma_0 \gamma_\mathrm{m} T^\mathrm{m}_{ii}
+ \sum_{\mathrm{m},\mathrm{n}} \gamma_\mathrm{m} \gamma_\mathrm{n} T^\mathrm{m}_{ij} T^\mathrm{n}_{ij}.
\label{eq:14}
\end{align}
Here, we introduce a fraction $f_\mathrm{m} (= \gamma_\mathrm{m}/\gamma_0)$ by assuming $\gamma_0 \neq 0$. Then,  Eq.~(\ref{eq:14}) reads
\begin{align}
2K = 
3 \gamma_0^2 \left( 1 + \frac{2}{3} \sum_\mathrm{m} f_\mathrm{m} T^\mathrm{m}_{ii}
+ \frac{1}{3} \sum_{\mathrm{m},\mathrm{n}} f_\mathrm{m} f_\mathrm{n} T^\mathrm{m}_{ij} T^\mathrm{n}_{ij} \right).
\label{eq:15}
\end{align}
Thus, $\gamma_0$ can be defined as
\begin{align}
\gamma_0 = \sqrt{\frac{2K}{3D}}, \ \
D = 1 + \frac{2}{3} \sum_\mathrm{m} f_\mathrm{m} T^\mathrm{m}_{ii}
+ \frac{1}{3} \sum_{\mathrm{m},\mathrm{n}} f_\mathrm{m} f_\mathrm{n} T^\mathrm{m}_{ij} T^\mathrm{n}_{ij}.
\label{eq:16}
\end{align}
Using this definition of $\gamma_0$, the model given by Eqs.~(\ref{eq:10}) and (\ref{eq:13}) always satisfies Eq.~(\ref{eq:14}). Note that $D > 0$ always holds because it is expressed by a square value, namely $D = ( \delta_{ij} + \sum_\mathrm{m} f_\mathrm{m} T^\mathrm{m}_{ij})^2/3$. Hence, the present modeling gives regular dimensional coefficients corresponding to $\nu_\mathrm{T}$ and $\zeta$'s, in contrast to the singular behavior of coefficients in EARSM \cite{gs1993,girimaji1996,akn1997,wj2000}. Although the present way to determine $\gamma_\mathrm{m}$'s is not a unique method for making the model satisfy Eq.~(\ref{eq:14}), it suggests a physically interesting feature of the model. To see this point clearly, we consider the following model:
\begin{align}
\sqrt{R}_{ij} = \gamma_0 \delta_{ij} - \gamma_S S_{ij} - \gamma_W W_{ij} - \gamma_N N_{ij},
\label{eq:17}
\end{align}
where $N_{ij}$ represents an additional model term. To construct a higher-order nonlinear eddy-viscosity model, we substitute $S_{ia} S_{ja}$ or $S_{ia} W_{aj} + S_{ja} W_{ai}$ into $N_{ij}$. To construct a more general model for the Reynolds stress, we substitute the helicity term \cite{yy1993,yb2016,inagakietal2017} or the time history effect of the mean strain tensor \cite{speziale1987,hd2008,ariki2019} into $N_{ij}$. It should be emphasized that the square root modeling makes the Reynolds stress always realizable, even when the complex model terms stated above are adopted to $N_{ij}$. The Reynolds stress given by Eqs.~(\ref{eq:10}), (\ref{eq:16}), and (\ref{eq:17}) is written as
\begin{align}
R_{ij} = \frac{2K}{3D} & \left[ \delta_{ij} - 2 f_S S_{ij} - f_N \left( N_{ij} + N_{ji} \right)
+ f_S^2 S_{ia} S_{aj}
- f_S f_W \left( S_{ia} W_{aj} + S_{ja} W_{ai} \right)
- f_W^2 W_{ia} W_{aj}
\right. \nonumber \\ 
& \left. + f_N f_S \left( N_{ia} S_{aj} + N_{ja} S_{ai} \right)
- f_N f_W \left( N_{ia} W_{aj} + N_{ja} W_{ai} \right)
+ f_N^2 N_{ia} N_{ja} \right],
\label{eq:18}
\end{align}
where
\begin{align}
D = 1 + \frac{1}{3} f_S^2 S^2 + \frac{1}{3} f_W^2 W^2 
- \frac{2}{3} f_N N_{ii}
+ \frac{2}{3} f_S f_N S_{ij} N_{ij}
+ \frac{2}{3} f_W f_N W_{ij} N_{ij}
+ \frac{1}{3} f_N^2 N^2,
\label{eq:19}
\end{align}
and $N^2 = N_{ij} N_{ij}$. To observe the physical property of the model more clearly, we adopt a simple coefficient form to $f_S$ and $f_W$; namely, we give $f_S = C_S K/\varepsilon$ and $f_W = C_W K/\varepsilon$, where $C_S$ and $C_W$ are constant. In such a case, the dimensional coefficient corresponds to the eddy viscosity $\nu_\mathrm{T}$ [see Eq.~(\ref{eq:8})] in the model given by Eqs.~(\ref{eq:18}) and (\ref{eq:19}) is written as follows:
\begin{align}
\nu_\mathrm{T} = \frac{4 K f_S}{3D} = \frac{4 C_S}{3D} \frac{K^2}{\varepsilon}
= \frac{4 C_S}{3} \frac{1}{1+ C_S^2 \hat{S}^2 /3 + C_W^2 \hat{W}^2 /3 + \cdots } \frac{K^2}{\varepsilon},
\label{eq:20}
\end{align}
where $\hat{S}^2 = K^2 S^2/\varepsilon^2$ and $\hat{W}^2 = K^2 W^2/\varepsilon^2$. Namely, the eddy viscosity is not simply expressed by Eq.~(\ref{eq:6}) but is incorporated with the effect of the mean strain rate and vorticity, $\hat{S}^2$ and $\hat{W}^2$, through its denominator. Equation~(\ref{eq:20}) suggests that the eddy viscosity depends on the mean strain or rotation rate. The mean strain rate- and vorticity-dependent expression of eddy viscosity is physically natural because $\nu_\mathrm{T}$ given by Eq.~(\ref{eq:6}) with a constant $C_\nu$ is not universal for several turbulent shear flows, as mentioned in Sec.~\ref{sec:level2a}. This functionalization of the dimensional coefficients leads to the regular behavior of the Reynolds stress for strongly strained or rotating turbulent flows. Moreover, the effect of the additional term $N_{ij}$ deductively enters the denominator of each term on the right-hand side of the Reynolds stress, as shown in Eqs.~(\ref{eq:18}) and (\ref{eq:19}). The present modeling provides a systematic way to determine a functional form of the dimensional coefficients for more general cases such as one that involves higher-order nonlinearity on the mean velocity gradient. 

The dependence of the eddy viscosity on the mean strain or rotation rate was analytically suggested by Okamoto \cite{okamoto1994} through TSDIA \cite{tsdia}. In their study, such an effect of the mean velocity gradient on the eddy viscosity appears as part of the cubic-nonlinear term. In the present modeling, it rather comes from the constraint given by Eqs.~(\ref{eq:15}) and (\ref{eq:16}). The mean strain rate- and vorticity-dependent form of the eddy viscosity was also discussed as the synthesized or composite time scale \cite{yoshizawaetal2006}. Hamba \cite{hamba2001} analytically obtained a quadratic-nonlinear realizable model through TSDIA \cite{tsdia}. In contrast to the theoretical approach by Hamba \cite{hamba2001}, we have to choose a set of terms for $\{ T^\mathrm{m}_{ij} \}$, such as $S_{ij}$ and $W_{ij}$, in the present modeling. However, this freedom of choice of $\{ T^\mathrm{m}_{ij} \}$ enables us to easily extend the realizable modeling to more general turbulent flows involving the turbulent helicity effect \cite{yy1993,yb2016,inagakietal2017} or the time-history effect of the mean strain rate \cite{speziale1987,hd2008,ariki2019}.

\section{\label{sec:level3}Construction of the model based on the square root technique}

\subsection{\label{sec:level3a}Quartic-nonlinear model of the Reynolds stress}

To construct a realizable model applicable to three-dimensional flows, we consider a simple quadratic-nonlinear form of the square root tensor on the velocity gradient:
\begin{gather}
\sqrt{R}_{ij} = \gamma_0 \delta_{ij} - \gamma_S S_{ij} - \gamma_W W_{ij} - \gamma_C C_{ij}, 
\label{eq:21} \\
C_{ij} = S_{ia} W_{aj} + S_{ja} W_{ai}.
\label{eq:22}
\end{gather}
Using Eq.~(\ref{eq:10}), the Reynolds stress is written as
\begin{align}
R_{ij} = \frac{2K}{3D} & \left[ \delta_{ij} - 2 f_S S_{ij} + f_S^2 S_{ia} S_{aj} 
- \left( 2 f_C + f_S f_W \right) C_{ij} - f_W^2 W_{ia} W_{aj} \right. \nonumber \\
& \left. 
+ f_C f_S \left( C_{ia} S_{aj} + C_{ja} S_{ai} \right)
- f_C f_W \left( C_{ia} W_{aj} + C_{ja} W_{ai} \right)
+ f_C^2 C_{ia} C_{aj} \right] ,
\label{eq:23}
\end{align}
where
\begin{align}
D = 1 + \frac{1}{3} f_S^2 S^2 + \frac{1}{3} f_W^2 W^2 + \frac{1}{3} f_C^2 C^2,
\label{eq:24}
\end{align}
and $C^2 = C_{ij} C_{ij}$. Note that $S_{ij} C_{ij} = W_{ij} C_{ij} = 0$. Because $C_{ij}$ is a quadratic-nonlinear term on the mean velocity gradient, the model given by Eq.~(\ref{eq:23}) corresponds to a quartic-nonlinear eddy-viscosity model. Although Eq.~(\ref{eq:23}) can be put into irreducible form through Cayley-Hamlitom's theorem as suggested by Pope \cite{pope1975}, we retain Eq.~(\ref{eq:23}) as the present form to simply express coefficients only by $f_S$, $f_W$, and $f_C$. The model given by Eq.~(\ref{eq:23}) comprises seven independent tensor bases except for $\delta_{ij}$. In the conventional turbulence modeling, there is no systematic way to determine the seven-dimensional coefficients, which makes the Reynolds stress satisfy the realizability. The square root modeling reduces the freedom of the dimensional coefficients by imposing the realizability as a guiding constraint. Consequently, we only need to determine three coefficients, $f_S$, $f_W$, and $f_C$, in the present modeling. 

To see some basic properties of the model given by Eq.~(\ref{eq:23}), we consider a unidirectional shear flow defined by $\bm{U} = (U_x, U_y, U_z)^\mathrm{t} = (U_x(y), 0,0)^\mathrm{t}$. In this flow, the Reynolds stress given by Eq.~(\ref{eq:23}) reads
\begin{subequations}
\begin{align}
R_{xx} & = \frac{2K}{3D} \left[ \left(1 + 2 f_C G^2\right)^2 + \left( f_S G + f_W G \right)^2 \right], 
\label{eq:25a} \\
R_{yy} & = \frac{2K}{3D} \left[ \left(1 - 2 f_C G^2\right)^2 + \left( f_S G - f_W G \right)^2 \right], 
\label{eq:25b} \\
R_{zz} & = \frac{2K}{3D}, 
\label{eq:25c} \\
R_{xy} & = - \frac{4K}{3D} \left( f_S - 2 f_W f_C G^2 \right) G, 
\label{eq:25d} \\
D & = 1 + \frac{2}{3} f_S^2 G^2 + \frac{2}{3} f_W^2 G^2 + \frac{8}{3} f_C^2 G^4,
\label{eq:25e}
\end{align}
\end{subequations}
where $G = \partial U_x/\partial y$ and $R_{xz} = R_{yz} = 0$. There are two points that should be noted. The first point is the anisotropy of the Reynolds stress. From many DNS and experiments of shear flows, such as a turbulent channel flow or a homogeneous turbulent shear flow, it is well-known that the diagonal components of the Reynolds stress show the inequality $R_{xx} > R_{zz} > R_{yy}$. However, the model given by Eqs.~(\ref{eq:25a})--(\ref{eq:25c}) results in $R_{zz} \le R_{yy}$ when $f_C = 0$. Hence, we demand $f_C > 0$. As seen from Eqs.~(\ref{eq:25a}) and (\ref{eq:25b}), $C_{ij}$ term plays the role of redistribution of the turbulence intensity between $R_{xx}$ and $R_{yy}$. When $f_C > 0$, the model given by Eqs.~(\ref{eq:25a})--(\ref{eq:25c}) can accurately predict the inequality for turbulent shear flows, $R_{xx} > R_{zz} > R_{yy}$.

The second point is the sign of the shear stress $R_{xy}$. If the shear rate increases and $2 f_W f_C G^2$ becomes larger than $f_S$, $f_S < 2 f_W f_C G^2$, the sign of $R_{xy}$ becomes the same as the mean velocity gradient $G$. This implies that the effective viscosity can be negative. The negative effective viscosity not only causes numerical instability but also gives a physically invalid condition for the unidirectional shear flow of the high-shear rate condition. Hence, we demand $f_S > 2 f_W f_C G^2$ for an arbitrary shear rate $G$ and adopt an appropriate expression for $f_W$ or $f_C$. 

Considering the above discussion, we propose a $K$-$\varepsilon$ model expression for the present model. Here, the following expressions for the coefficients $f_S$, $f_W$, and $f_C$ are adopted:
\begin{subequations}
\begin{gather}
f_S = f_W = C_1 f_\nu \frac{K}{\varepsilon}
\label{eq:26a} \\
f_C = \frac{C_2}{1 + C_3 (\hat{S}^2 + \hat{W}^2)} \frac{K^2}{\varepsilon^2},
\label{eq:26b}
\end{gather}
\end{subequations}
where $f_\nu$ is a damping function introduced for the connection of the model to the solid wall and $\hat{S}^2$ and $\hat{W}^2$ are already defined in connection to Eq.~(\ref{eq:20}). $C_1$, $C_2$, and $C_3$ are positive model constants. Because $f_C$ depends on the mean strain and rotation rate, it can be calibrated to satisfy $f_S > 2 f_W f_C G^2$ for a high-shear-rate condition of the unidirectional shear flow. Of course, Eqs.~(\ref{eq:26a}) and (\ref{eq:26b}) are not a unique choice of $f_S$, $f_W$, and $f_C$, but are a simple form that avoids the irregular condition for the shear stress $R_{xy}$. In addition, we do not pay much attention to the behavior of the model around TCL condition in the present study. To consider TCL condition of ARSM in the vicinity of the solid wall, we may have to put additional terms expressing the near wall effect \cite{ajl2003}. Such a modeling of TCL physics is a future study. In the present study, we rather focus on the bulk behavior away from the solid wall.

\subsection{\label{sec:level3b}Anisotropy tensor in unidirectional shear flow}

\begin{figure}[htp]
\centering
\includegraphics[scale=0.65]{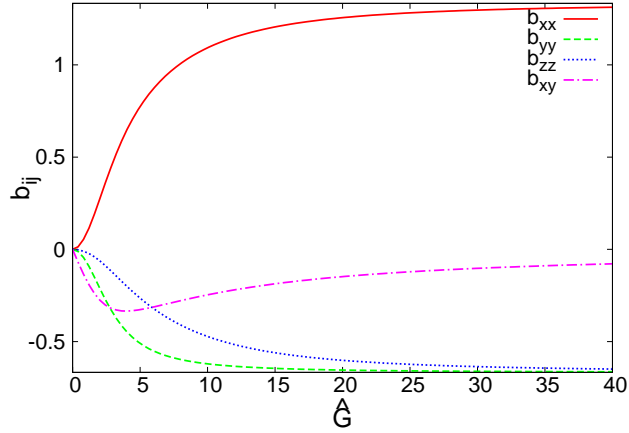}
\caption{Profile of the anisotropy tensor of the Reynolds stress $b_{ij}$ against the non-dimensional shear rate $\hat{G}$.}
\label{fig:1}
\end{figure}

We examine the anisotropic behavior of the model given by Eqs.~(\ref{eq:25a})--(\ref{eq:25e}), (\ref{eq:26a}), and (\ref{eq:26b}) for a high-shear-rate case in the unidirectional shear flow. The anisotropy tensor of the Reynolds stress $b_{ij}$ is defined as
\begin{align}
b_{ij} = \frac{R_{ij}}{K} - \frac{2}{3} \delta_{ij}.
\label{eq:27}
\end{align}
The anisotropy tensor for the model given by Eqs.~(\ref{eq:25a})--(\ref{eq:25e}), (\ref{eq:26a}), and (\ref{eq:26b}) is written as
\begin{subequations}
\begin{align}
b_{xx} & = \frac{2}{3D} \left[ \left(1 + \frac{2C_2}{1+4C_3 \hat{G}^2} \hat{G}^2\right)^2 + \left( 2 C_1 \hat{G} \right)^2 \right]
-\frac{2}{3}, 
\label{eq:28a} \\
b_{yy} & = \frac{2}{3D} \left(1 - \frac{2C_2}{1+4C_3 \hat{G}^2} \hat{G}^2\right)^2 -\frac{2}{3}, 
\label{eq:28b} \\
b_{zz} & = \frac{2}{3D} -\frac{2}{3}, 
\label{eq:28c} \\
b_{xy} & = - \frac{4C_1}{3D} \left( 1 - \frac{2C_2}{1+4C_3 \hat{G}^2} \hat{G}^2 \right) \hat{G}, 
\label{eq:28d} \\
D & = 1 + \frac{4}{3} C_1^2 \hat{G}^2 + \frac{8}{3} \left( \frac{C_2}{1+4C_3 \hat{G}^2} \right)^2 \hat{G}^4,
\label{eq:28e}
\end{align}
\end{subequations}
where $\hat{G} = KG/\varepsilon$ and $b_{xz} = b_{yz} = 0$. Note that the damping function $f_\nu$ is not concerned here; namely, $f_\nu = 1$. In this case, $b_{ij}$ is determined only by the non-dimensional shear rate $\hat{G}$. It is clearly seen that the condition $b_{zz} > b_{yy}$ is satisfied for finite $\hat{G}$ when $C_2/(2C_3) < 1$. Moreover, the sign of $b_{xy}$ is always opposite to that of $\hat{G}$ as long as $C_2/(2C_3) < 1$. Figure~\ref{fig:1} shows the profile of the anisotropy tensor against the non-dimensional shear rate $\hat{G}$. Here, $C_1 = 0.13$, $C_2 = 0.021$, and $C_3 = 0.018$ are adopted. The first equation of the realizability condition given by Eq.~(\ref{eq:1a}) is rewritten as
\begin{align}
-\frac{2}{3} \le b_{\alpha \alpha} \le \frac{4}{3}.
\label{eq:29}
\end{align}
It is confirmed that the present model exactly satisfies Eq.~(\ref{eq:29}) and reproduces the physical features in shear flows, namely $b_{xx} > b_{zz} > b_{yy}$ and $b_{xy} < 0$, even when the non-dimensional shear rate $\hat{G}$ is large.

\subsection{\label{sec:level3c}Shear stress in rotating shear flow}

\begin{figure}[htp]
\centering
\includegraphics[scale=0.65]{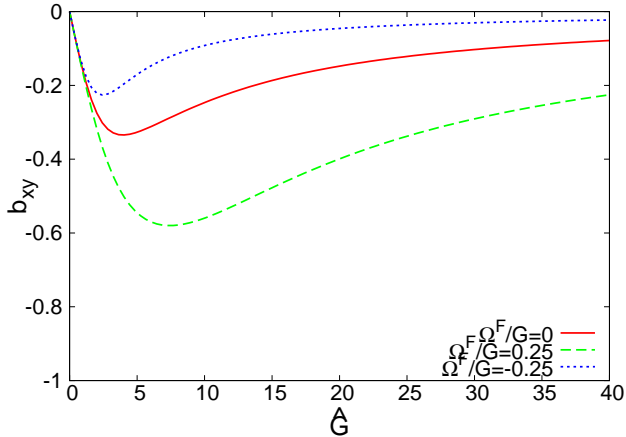}
\caption{Profiles of $b_{xy}$ against the non-dimensional shear rate $\hat{G}$ for $\Omega^\mathrm{F}/G = 0, 0.25$, and $-0.25$ in the spanwise rotating shear flow.}
\label{fig:2}
\end{figure}

For another simple shear flow case, we consider a spanwise rotating shear flow \cite{bardinaetal1983}. We consider the same unidirectional shear flow as previous sections and set the rotation axis in the $z$ direction, $\bm{\Omega}^\mathrm{F} = (0,0,\Omega^\mathrm{F})^\mathrm{t}$ where $\bm{\Omega}^\mathrm{F}$ denotes the angular velocity of frame rotation. In this condition, $b_{xy}$ is written as
\begin{subequations}
\begin{align}
b_{xy} & = - \frac{4C_1}{3D} \left\{ 1 - \frac{2C_2}{1+2C_3 \hat{G}^2 [1 + (1 - 2 \Omega^\mathrm{F}/G)^2] } \hat{G}^2 (1 - 2 \Omega^\mathrm{F}/G)^2 \right\} \hat{G}, 
\label{eq:30a} \\
D & = 1 + \frac{2}{3} C_1^2 \hat{G}^2 [1 + (1 - 2 \Omega^\mathrm{F}/G)^2]
+ \frac{8}{3} \left\{ \frac{C_2}{1+2C_3 \hat{G}^2 [1 + (1 - 2 \Omega^\mathrm{F}/G)^2]} \right\}^2 \hat{G}^4 (1 - 2 \Omega^\mathrm{F}/G)^2.
\label{eq:30b}
\end{align}
\end{subequations}
Here, a new parameter, $\Omega^\mathrm{F}/G$, appears compared with Eqs.~(\ref{eq:28d}) and (\ref{eq:28e}). Of course, Eqs.~(\ref{eq:30a}) and (\ref{eq:30b}) reduce to Eqs.~(\ref{eq:28d}) and (\ref{eq:28e}) when $\Omega^\mathrm{F}/G = 0$. Figure~\ref{fig:2} shows the profiles of $b_{xy}$ against the non-dimensional shear rate $\hat{G}$ for $\Omega^\mathrm{F}/G = 0, 0.25$, and $-0.25$. When the rotation is against the shear, $\Omega^\mathrm{F}/G = 0.25$, the shear stress $b_{xy}$ decreases, while it increases when the rotation is with the shear, $\Omega^\mathrm{F}/G = -0.25$. This result is consistent with the large-eddy simulation performed by Bardina \textit{et al}. \cite{bardinaetal1983}.

\section{\label{sec:level4}Numerical verification}

In this section, we numerically verify the performance of the present model in fundamental turbulent shear flows. First, we perform a turbulent channel flow as a basic wall shear flow. Second, we perform a homogeneous turbulent shear flow as an example where the linear eddy-viscosity model given by Eqs.~(\ref{eq:4}) and (\ref{eq:6}) fails to predict the evolution of the turbulent kinetic energy. Third, we perform an axially rotating turbulent pipe flow as an example of the three-dimensional flow. In this flow, no quadratic-nonlinear eddy-viscosity model can predict the mean tangential or swirl velocity \cite{syb2000}. In contrast to the conventional realizable model, such as that given by Shih \textit{et al}. \cite{szl1993,szl1995}, the present model involves cubic nonlinearity on the velocity gradient, as seen in Eq.~(\ref{eq:23}), and thus, is able to predict the mean tangential or swirl velocity.

\subsection{\label{sec:level4a}Turbulent channel flow}

In the turbulent channel flow, the governing equations for the $K$-$\varepsilon$ model are given by
\begin{align}
\frac{\partial U_x}{\partial t} & = 
-\frac{\partial R_{xy}}{\partial y} + \nu \frac{\partial^2 U_x}{\partial y^2} + 1,
\label{eq:31} \\
\frac{\partial K}{\partial t} & =
- R_{xy} \frac{\partial U_x}{\partial y} - \varepsilon
+\frac{\partial}{\partial y} \left[ \left( \frac{\nu_\mathrm{T}}{\sigma_K} + \nu \right)
\frac{\partial K}{\partial y} \right],
\label{eq:32} \\
\frac{\partial \varepsilon}{\partial t} & =
- C_{\varepsilon 1} \frac{\varepsilon}{K} R_{xy} \frac{\partial U_x}{\partial y} 
- C_{\varepsilon 2} f_\varepsilon \frac{\varepsilon^2}{K}
+\frac{\partial}{\partial y} \left[ \left( \frac{\nu_\mathrm{T}}{\sigma_\varepsilon} + \nu \right)
\frac{\partial \varepsilon}{\partial y} \right],
\label{eq:33}
\end{align}
where $x$ and $y$ denote the streamwise and wall-normal directions, respectively. The length and velocity are normalized by the channel half width $h$ and the wall friction velocity $u_\tau (= \sqrt{ \nu |\partial U_x/\partial y|_\text{wall}|} )$, respectively. $\sigma_K$, $C_{\varepsilon 1}$, $C_{\varepsilon 2}$, and $\sigma_\varepsilon$ are the model constants. $f_\varepsilon$ denotes the damping function for the destruction rate of the dissipation rate $\varepsilon$. The performance of the present model is compared with that of the linear eddy-viscosity model, in which the Reynolds stress is given by Eq.~(\ref{eq:4}). For the linear eddy-viscosity model, which is applicable to the non-slip boundary condition, we use that given by Abe \textit{et al}. \cite{akn1994} (hereafter the AKN model). The results are compared with those of the DNS performed by Moser \textit{et al}. \cite{mkm1999} at $\mathrm{Re}_\tau (= u_\tau h/\nu) = 590$. We adopt the same damping functions for the present model as those for the AKN model, which are described by the distance from the wall normalized by the Kolmogorov length scale $\eta [= (\nu^3/ \varepsilon)^{1/4}]$ and the turbulent Reynolds number $\mathrm{Re}_\mathrm{T} (=K^2/\nu \varepsilon)$:
\begin{align}
f_\nu & = \left\{ 1 -\mathrm{exp} \left[ - \frac{y}{a_1\eta} \right] \right\}^2
\left\{ 1 + \frac{a_2}{\mathrm{Re}_\mathrm{T}^{3/4}} \mathrm{exp} \left[ - \left( \frac{\mathrm{Re}_\mathrm{T}}{a_3} \right)^2 \right] \right\}, 
\label{eq:34} \\
f_\varepsilon & = \left\{ 1 -\mathrm{exp} \left[ - \frac{y}{a_{\varepsilon 1} \eta} \right] \right\}^2
\left\{ 1 - a_{\varepsilon 2} \mathrm{exp} \left[ - \left( \frac{\mathrm{Re}_\mathrm{T}}{a_{\varepsilon 3}} \right)^2 \right] \right\}.
\label{eq:35}
\end{align}
An advantage of using the Kolmogorov length scale for the damping functions is that they can have a non-zero value even when the mean velocity gradient at the wall is zero, $\partial U_x/\partial y|_\text{wall} = 0$, while the conventional damping functions described by the distance from the wall normalized by $\nu/u_\tau$ gives $f_\nu = 0$. The eddy viscosity $\nu_\mathrm{T}$ in Eqs.~(\ref{eq:32}) and (\ref{eq:33}) is given by
\begin{align}
\nu_\mathrm{T} = C_\nu f_\nu \frac{K^2}{\varepsilon}.
\label{eq:36}
\end{align}
For the linear eddy-viscosity model given by Eq.~(\ref{eq:4}), we use the same expression as Eq.~(\ref{eq:36}). In the present model, the same model constants as those in the AKN model are adopted and these values are given in Table \ref{tb:1}. For the unidirectional shear flows, the only difference between the present and AKN models is the model of the Reynolds shear stress $R_{xy}$; in the present model, $R_{xy}$ is given by Eqs.~(\ref{eq:25d}), (\ref{eq:26a}), and (\ref{eq:26b}), while in the AKN model, it is given by
\begin{align}
R_{xy} = - \nu_\mathrm{T} \frac{\partial U_x}{\partial y} = - C_\nu f_\nu \frac{K^2}{\varepsilon} G.
\label{eq:37}
\end{align}
Owing to the damping function $f_\nu$, both the present and AKN models achieve the near wall asymptotic condition for the Reynolds shear stress, $R_{xy} \sim y^3$.

\begin{table}[t]
\centering
\caption{Model constants.}
\begin{ruledtabular}
\begin{tabular}{ccccccccccccccc}
Model & $C_1$ & $C_2$ & $C_3$ & $C_\nu$ & $C_{\varepsilon 1}$ & $C_{\varepsilon 2}$ & $\sigma_K$ & $\sigma_\varepsilon$ 
& $a_1$ & $a_2$ & $a_3$ & $a_{\varepsilon 1}$ & $a_{\varepsilon 2}$ & $a_{\varepsilon 3}$ \\ \hline
present & 0.13 & 0.021 & 0.018 & 0.09 & 1.5 & 1.9 & 1.4 & 1.4 & 14 & 5 & 200 & 3.1 & 0.3 & 6.5 \\
linear \cite{akn1994} & - & - & - & 0.09 & 1.5 & 1.9 & 1.4 & 1.4 & 14 & 5 & 200 & 3.1 & 0.3 & 6.5
\end{tabular}
\end{ruledtabular}
\label{tb:1}
\end{table}

\begin{figure}[htp]
\centering
\includegraphics[scale=0.65]{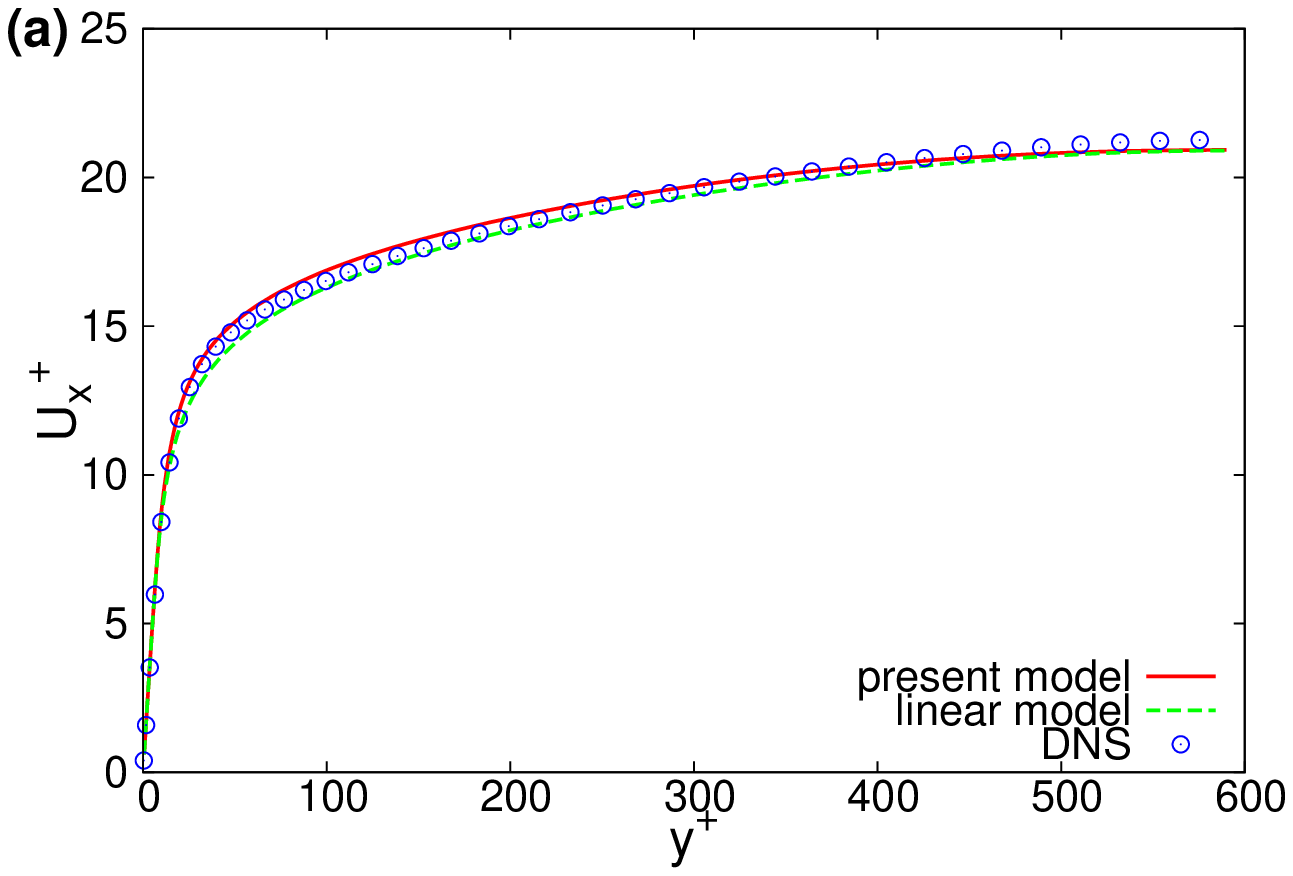}
 \begin{tabular}{c}
  \begin{minipage}{0.49\hsize}
   \centering
   \includegraphics[scale=0.65]{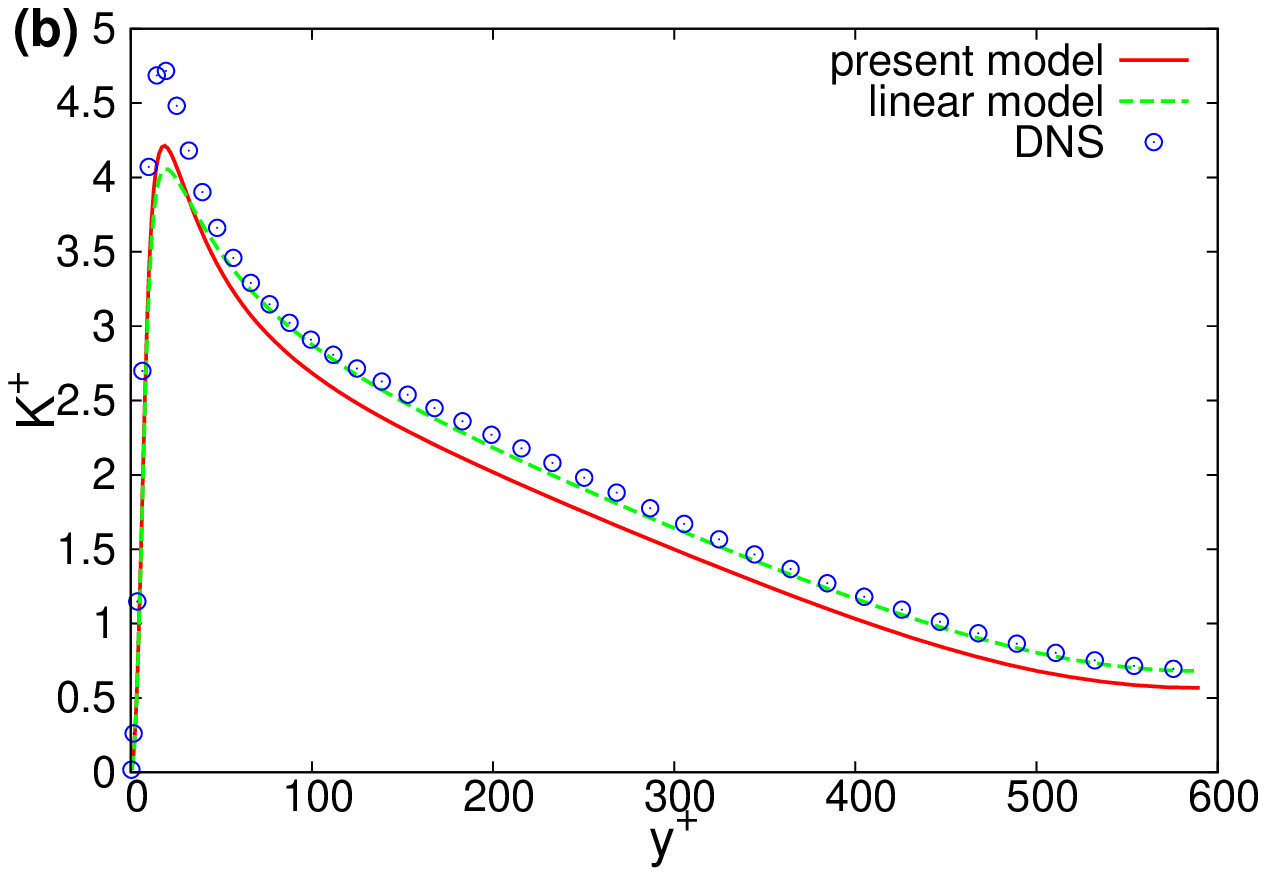}
  \end{minipage}

  \begin{minipage}{0.49\hsize}
   \centering
   \includegraphics[scale=0.65]{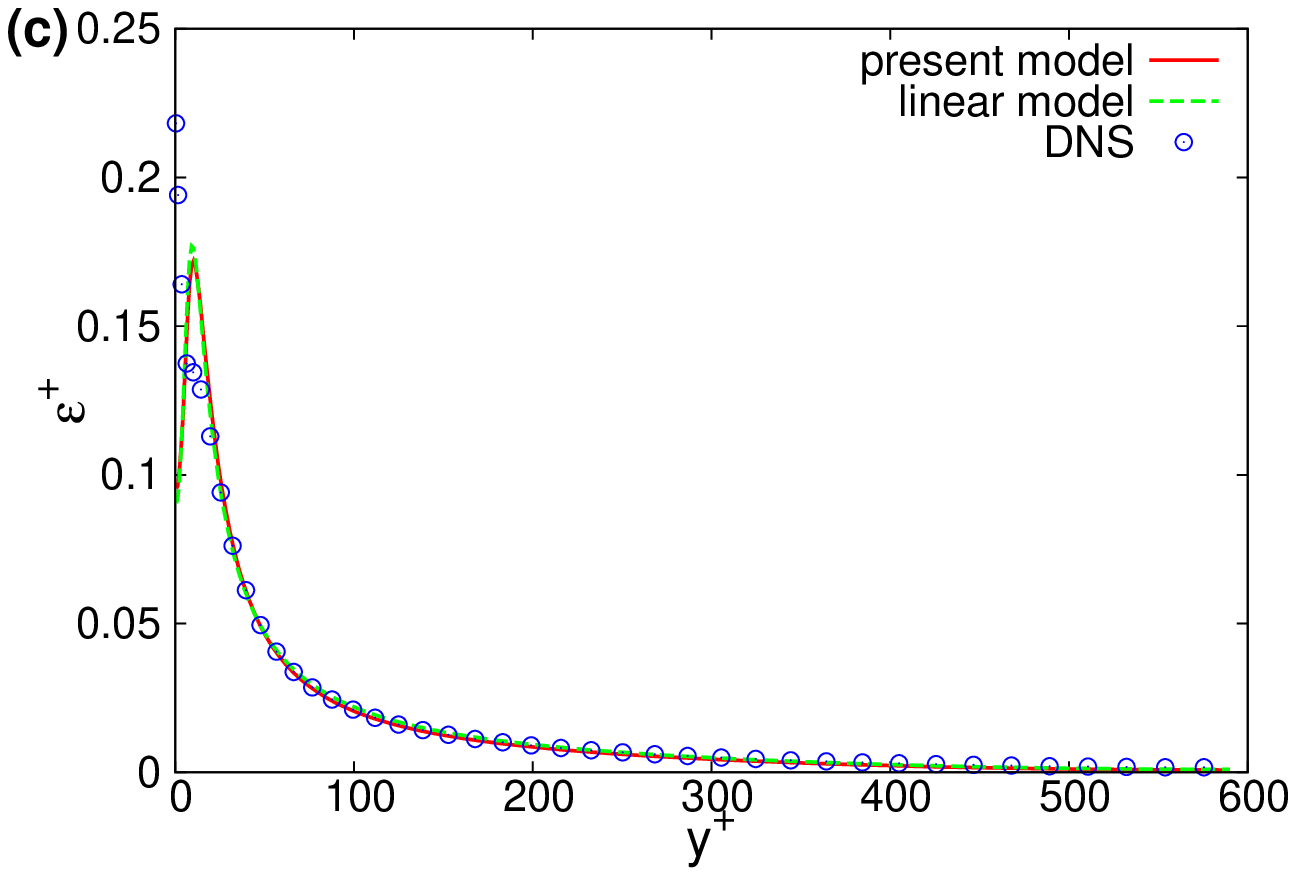}
  \end{minipage}
 \end{tabular}
\caption{Profiles of (a) the mean velocity $U_x^+$, (b) the turbulent energy $K^+$, and (c) the dissipation rate $\varepsilon^+$ in the turbulent channel flow at $\mathrm{Re}_\tau = 590$. Here, $y^+ = y u_\tau/\nu$, $U_x^+ = U_x/u_\tau$, $K^+ = K/u_\tau^2$, and $\varepsilon^+ = \varepsilon \nu /u_\tau^4$.}
\label{fig:3}
\end{figure}

The profile of the mean velocity is shown in Fig.~\ref{fig:3}(a). Both the present and linear models give a good result compared to DNS. Figure~\ref{fig:3}(b) and (c) show the profiles of the turbulent energy and its dissipation rate, respectively. The prediction of the present model is almost similar to that of the linear eddy-viscosity model. In the near wall region at $y^+ < 20$, both the present and linear models slightly underestimate the turbulent energy $K$ and overestimate  the dissipation rate $\varepsilon$. In this sense, some modification is required for predicting the near wall behavior. Overall, however, the present model gives a good prediction for the basic properties in the turbulent channel flow. Note that the damping function works only at $y^+ < 100$, so that the behavior of outer region, $y^+ >100$, is independent of the damping function.

\begin{figure}[htp]
\centering
 \begin{tabular}{c}
  \begin{minipage}{0.49\hsize}
   \centering
   \includegraphics[scale=0.65]{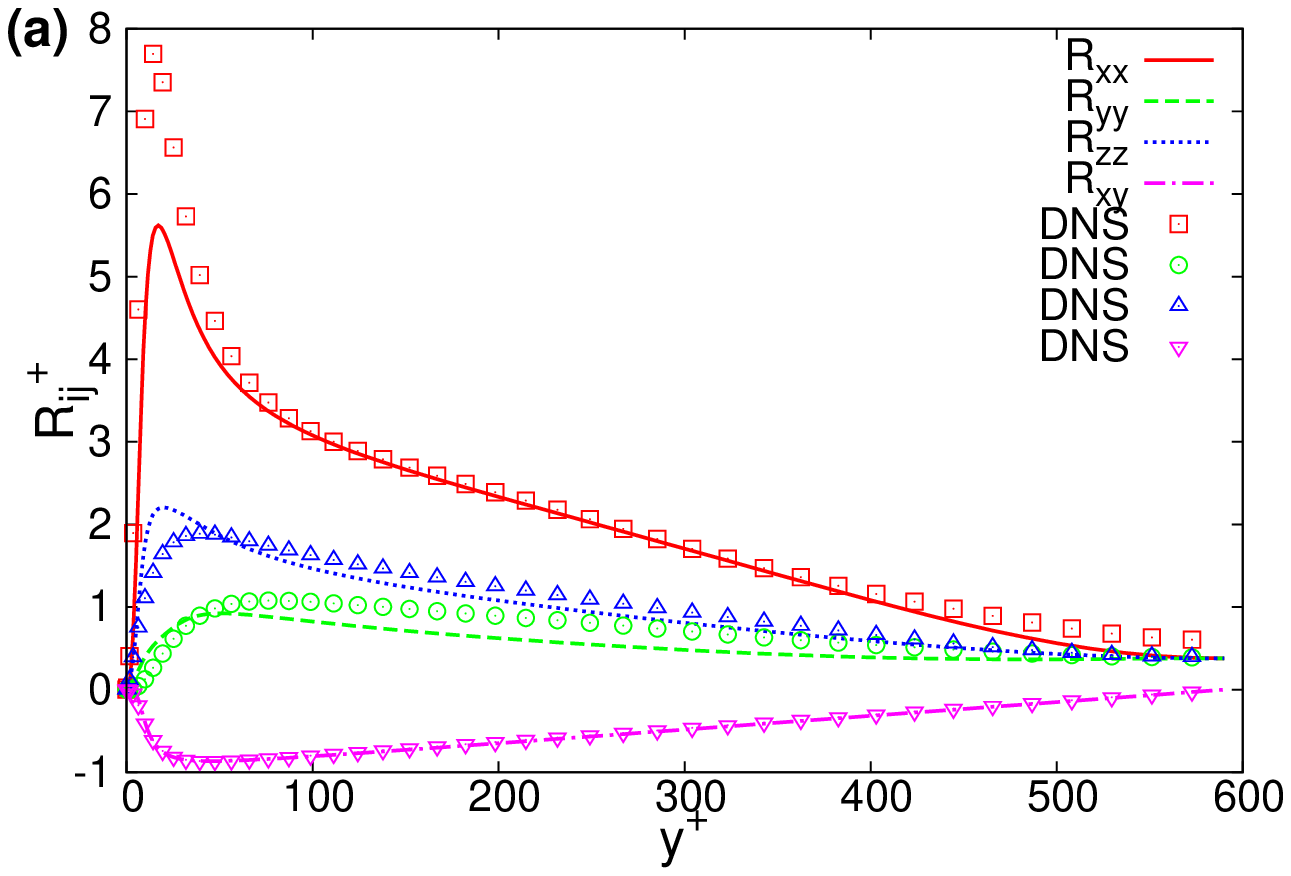}
  \end{minipage}
  \begin{minipage}{0.49\hsize}
   \centering
   \includegraphics[scale=0.65]{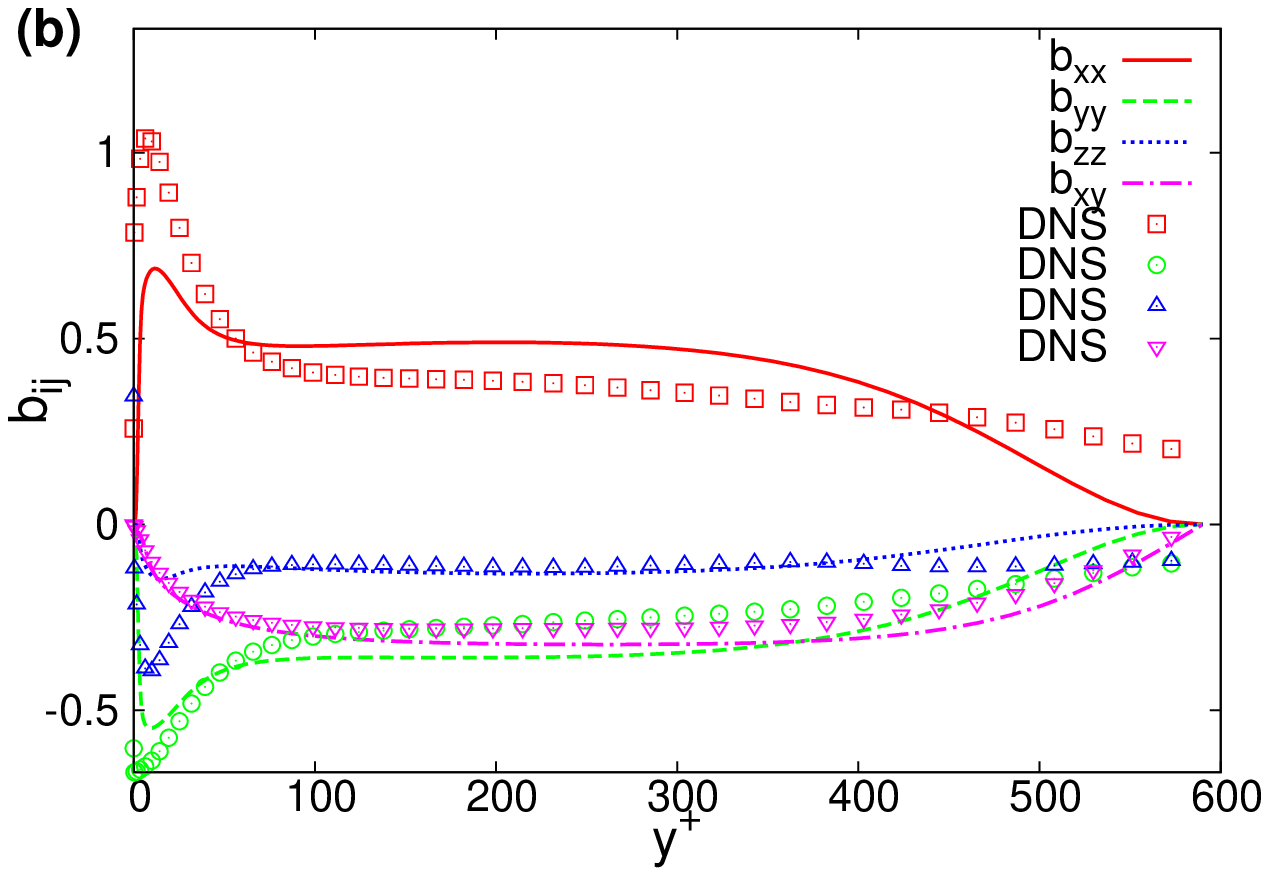}
  \end{minipage}
 \end{tabular}
\caption{Profiles of (a) the Reynolds stress and (b) the anisotropy tensor of the Reynolds stress for the present model in the turbulent channel flow at $\mathrm{Re}_\tau = 590$.}
\label{fig:4}
\end{figure}

\begin{figure}[htp]
\centering
\includegraphics[scale=0.65]{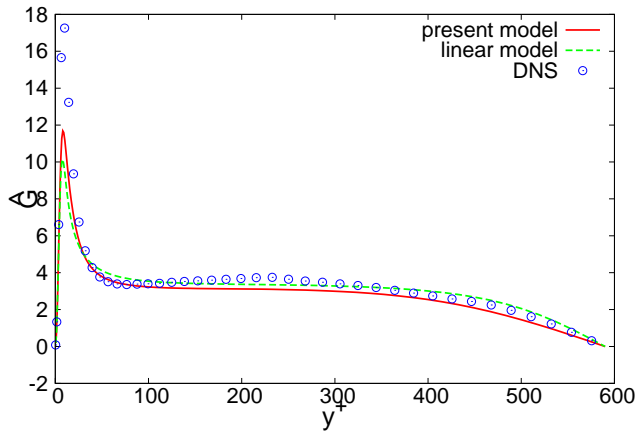}
\caption{Profile of the non-dimensional shear rate $\hat{G}$ in the turbulent channel flow at $\mathrm{Re}_\tau = 590$.}
\label{fig:5}
\end{figure}

In the RANS model of the unidirectional shear flows, the diagonal components of the Reynolds stress do not appear in the governing equations. In a turbulent flow in a square duct, however, the anisotropic property of the Reynolds stress is essential for predicting the secondary flow \cite{speziale1987}. In this sense, it is worth examining the anisotropic property of the model for a wall-bounded turbulent flow. It is well-known that the linear eddy-viscosity model given by Eq.~(\ref{eq:4}) cannot predict the anisotropy of the Reynolds stress in the unidirectional shear flow; namely, the linear eddy-viscosity model leads to
\begin{align}
R_{xx} = R_{yy} = R_{zz} =\frac{2}{3} K,
\label{eq:38}
\end{align}
while the experiments and DNS show $R_{xx} > R_{zz} > R_{yy}$ in the turbulent channel flow. As discussed in Sec.~\ref{sec:level3}, the present model can predict this inequality. The profile of each component of the Reynolds stress for the present model is shown in Fig.~\ref{fig:4}(a). The profile of the anisotropy tensor of the Reynolds stress $b_{ij}$ is also shown in Fig.~\ref{fig:4}(b). It is seen that the present model well-predicts the anisotropic property at $100 < y^+ < 400$. In the near wall region at $y^+ < 50$, the present model underestimates the streamwise component of the Reynolds stress, as seen in Fig.~\ref{fig:4}(a) and (b). This is mainly caused by the failure of prediction of the non-dimensional shear rate $\hat{G}$. The profile of $\hat{G}$ is shown in Fig.~\ref{fig:5}. It is seen that $\hat{G}$ at $y^+ \sim 10$ is about two thirds the DNS for the present model. This tendency is the same as that of the linear model. This result comes from the underestimation of $K$ and the overestimation of $\varepsilon$ at $y^+ \sim 10$, because the mean velocity is well-predicted by both the present and linear models. Hence, we need to modify the transport equations for $K$ and $\varepsilon$ at the near wall region to improve the accuracy of the model. However, we do not treat this point further.

\subsection{\label{sec:level4b}Homogeneous turbulent shear flow}

As discussed in Sec.~\ref{sec:level2a}, the standard type of the linear eddy-viscosity model given by Eqs.~(\ref{eq:4}) and (\ref{eq:6}) overestimates the time evolution of the turbulent energy in a homogeneous turbulent shear flow. As the conventional ARSM  overcomes this shortfall with the aid of the mean strain or rotation rate dependence of the dimensional coefficients \cite{taulbee1992,gs1993,girimaji1996,akn1997,szl1993,szl1995,hamba2001,yoshizawaetal2006}, the present model also does. The governing equations for the $K$-$\varepsilon$ model are given by Eqs.~(\ref{eq:32}) and (\ref{eq:33}), where the diffusion terms, the third term in them, vanish and $f_\nu = f_\varepsilon = 1$ in the homogeneous turbulent shear flow. The reference DNS was performed by Hamba \cite{hamba2001}, where the mean shear rate was set to $G = 7.67$, the initial value of the Taylor microscale Reynolds number $\mathrm{Re}_\lambda (= \sqrt{20K^2/3\nu \varepsilon})$ was set to $\mathrm{Re}_\lambda = 24$, the initial value of the turbulent energy was set to $K = 1/2$, and the initial energy spectrum was proportional to $k^4 \exp [-2k^2/k_\mathrm{p}^2]$, where $k_\mathrm{p} = 9$. As the initial conditions for the $K$-$\varepsilon$ model, the DNS values of $K$ and $\varepsilon$ at time $Gt=6$ are adopted, where the energy spectrum is fully developed in the high-wavenumber region. 

\begin{figure}[htp]
\centering
 \begin{tabular}{c}
  \begin{minipage}{0.49\hsize}
   \centering
   \includegraphics[scale=0.65]{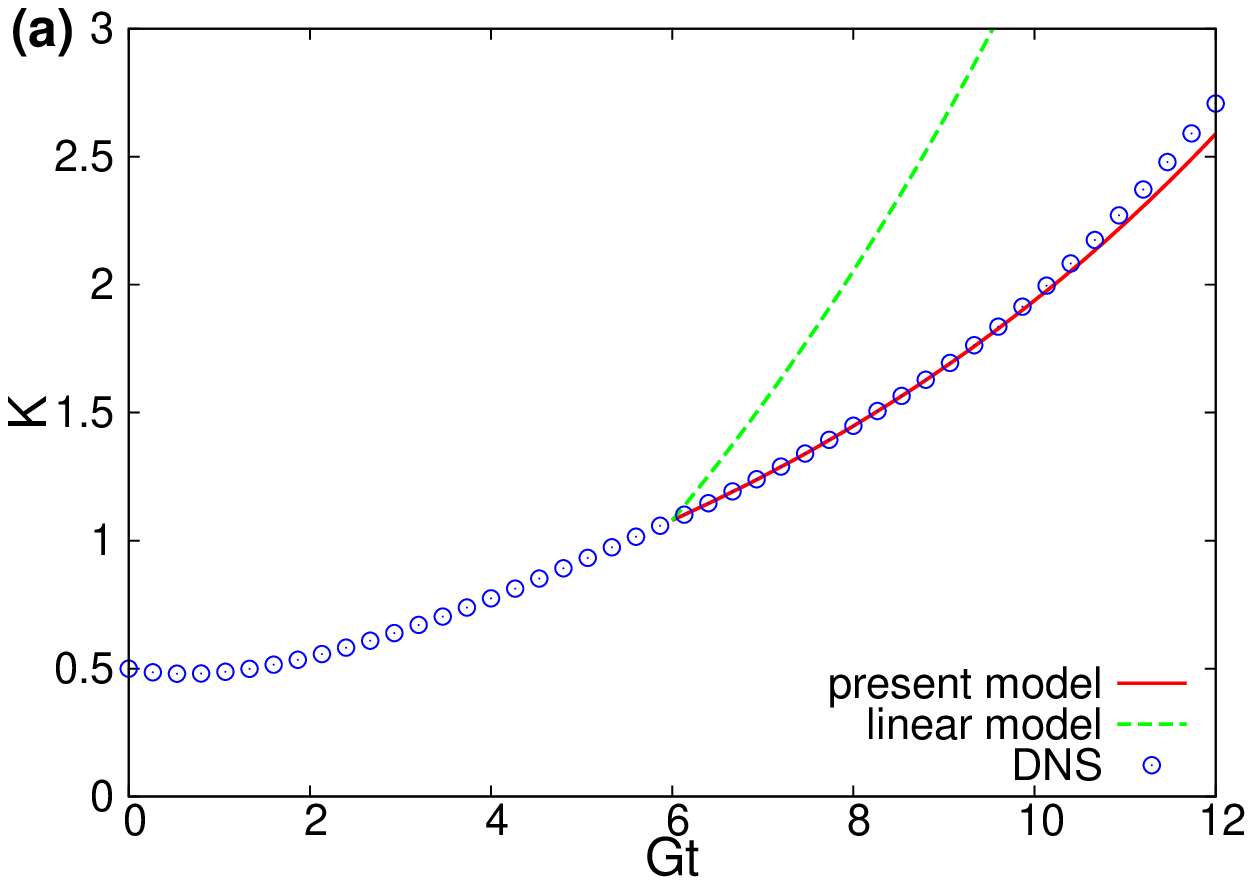}
  \end{minipage}

  \begin{minipage}{0.49\hsize}
   \centering
   \includegraphics[scale=0.65]{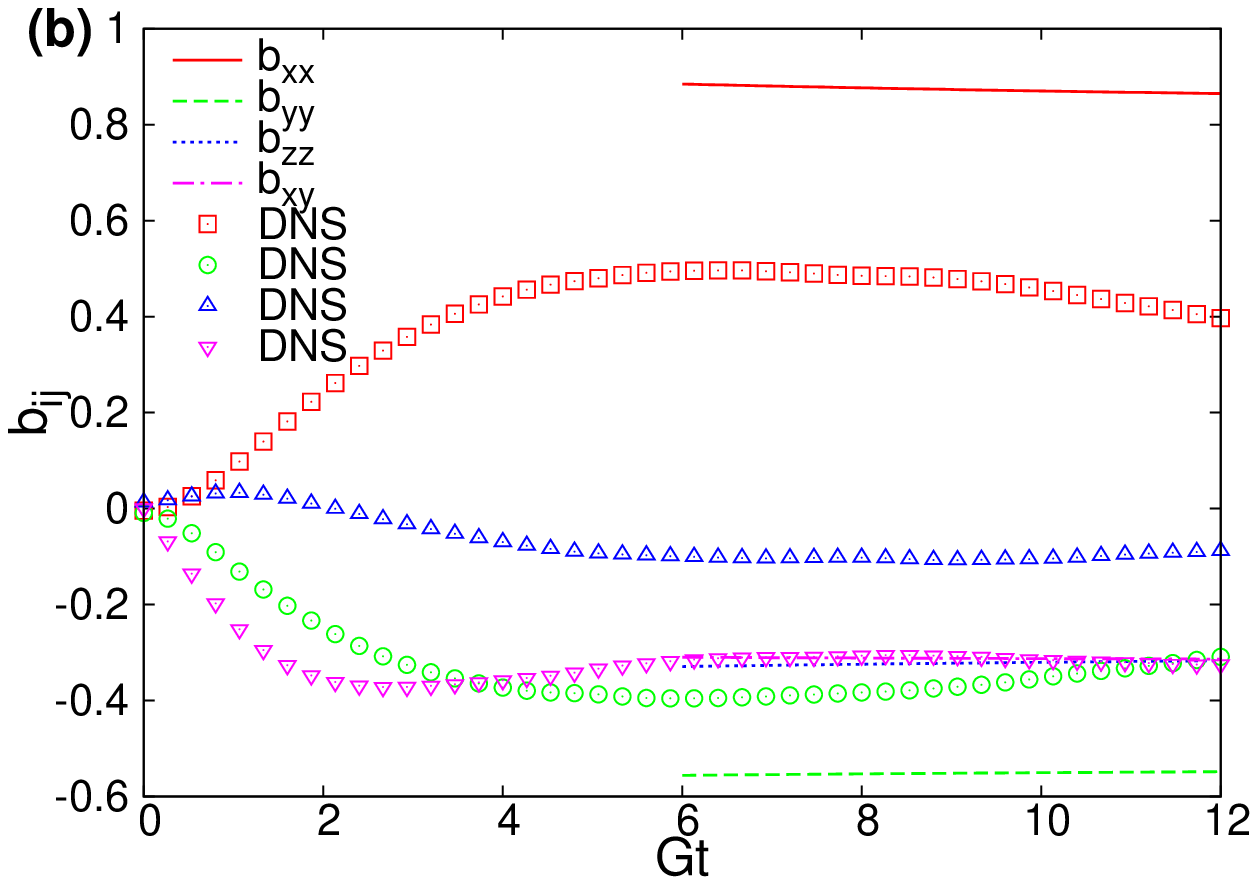}
  \end{minipage}
 \end{tabular}
\caption{Time evolution of (a) the turbulent energy and (b) the anisotropy tensor of the Reynolds stress in the homogeneous turbulent shear flow.}
\label{fig:6}
\end{figure}

The time evolution of the turbulent energy is shown in Fig.~\ref{fig:6}(a), which also plots the result of the linear eddy-viscosity model for reference, where the set of the model constants is the same as that for the AKN model \cite{akn1994} given in Table~\ref{tb:1}. As seen in Fig.~\ref{fig:6}(a), the present model well-predicts the time evolution of the turbulent energy, while the linear model overestimates it. The time evolution of the anisotropy tensor of the Reynolds stress is shown in Fig,~\ref{fig:6}(b). The good agreement of $b_{xy}$ with the DNS value provides the good prediction of the turbulent energy. However, the diagonal components of the anisotropy tensor are somewhat overestimated. This result indicates that the present model tends to overestimate the anisotropy for high-shear-rate turbulent flows. 

In the present study, we construct a model for $\sqrt{R}_{ij}$ in terms of $S_{ij}$, $W_{ij}$, and $C_{ij}$ for simplicity of the model expression. Hence, there is still enough room for further development, e.g., introduction of additional terms such as $S_{ia}S_{aj}$, $W_{ia}W_{aj}$, or the time history effect of the mean strain tensor \cite{speziale1987,hd2008,ariki2019} to $\sqrt{R}_{ij}$, or modifications of $f_S$, $f_W$, and $f_C$. Namely, the overestimation of anisotropy in the homogeneous turbulent shear flow can be improved by more sophisticated modeling.

\subsection{\label{sec:level4c}Axially rotating turbulent pipe flow}

A prominent feature of the present model is that it involves cubic nonlinearity on the mean velocity gradient, which is an essential for predicting the mean swirl flow in an axially rotating turbulent pipe flow \cite{syb2000}. This point is a critical difference between the present and the conventional realizable models \cite{szl1993,szl1995,akn1997,hamba2001}. Assuming the homogeneity of the turbulence field in the axial and azimuthal directions, the mean velocity can be written as $\bm{U} = (U_r, U_\theta, U_z) = (0,U_\theta (r), U_z(r))$ in the cylindrical coordinates. The governing equations for the $K$-$\varepsilon$ model in an inertial frame of the cylindrical coordinates are written as follows:
\begin{align}
\frac{\partial U_\theta}{\partial t} & = 
-\frac{1}{r} \frac{\partial}{\partial r} (r R_{r \theta}) - \frac{R_{r \theta}}{r}
+ \nu \left[ \frac{1}{r} \frac{\partial}{\partial r} \left( r \frac{\partial U_\theta}{\partial r} \right) - \frac{U_\theta}{r^2} \right],
\label{eq:39} \\
\frac{\partial U_z}{\partial t} & = 
-\frac{1}{r} \frac{\partial}{\partial r} (r R_{rz})
+ \nu \frac{1}{r} \frac{\partial}{\partial r} \left( r \frac{\partial U_z}{\partial r} \right) + f^\mathrm{ex},
\label{eq:40} \\
\frac{\partial K}{\partial t} & =
- S_{r \theta} R_{r \theta} - S_{rz} R_{rz} - \varepsilon
+\frac{1}{r} \frac{\partial}{\partial r} \left[ \left( \frac{\nu_\mathrm{T}}{\sigma_K} + \nu \right)
r \frac{\partial K}{\partial r} \right],
\label{eq:41} \\
\frac{\partial \varepsilon}{\partial t} & =
- C_{\varepsilon 1} \frac{\varepsilon}{K} \left( S_{r \theta} R_{r \theta} + S_{rz} R_{rz} \right)
- C_{\varepsilon 2} f_\varepsilon \frac{\varepsilon^2}{K}
+\frac{1}{r} \frac{\partial}{\partial r} \left[ \left( \frac{\nu_\mathrm{T}}{\sigma_\varepsilon} + \nu \right)
r \frac{\partial \varepsilon}{\partial r} \right] .
\label{eq:42}
\end{align}
Here, the velocity and length scale are normalized by the bulk mean velocity $U_\mathrm{m} [=(2/R^2) \int_0^R \mathrm{d}r \ r U_z]$ and the pipe radius $R$, respectively. $f^\mathrm{ex}$ denotes the external forcing that keeps the bulk Reynolds number $\mathrm{Re}_\mathrm{m} (= U_\mathrm{m} R/\nu)$ constant. In the cylindrical coordinates, we can write a second-rank tensor $\bm{\mathsf{T}}$ as
\begin{align}
\bm{\mathsf{T}} =
\begin{bmatrix}
T_{rr} & T_{r\theta} & T_{rz} \\
T_{\theta r} & T_{\theta \theta} & T_{\theta z} \\
T_{zr} & T_{z \theta} & T_{zz}
\label{eq:43}
\end{bmatrix}.
\end{align}
In an axially rotating turbulent pipe flow, the mean strain rate and mean absolute vorticity tensors are respectively written as follows:
\begin{align}
\bm{\mathsf{S}} =
\begin{bmatrix}
0 & \displaystyle r \frac{\partial}{\partial r} \left( \frac{U_\theta}{r} \right) & \displaystyle \frac{\partial U_z}{\partial r} \\
\displaystyle r \frac{\partial}{\partial r} \left( \frac{U_\theta}{r} \right) & 0 & 0 \\
\displaystyle \frac{\partial U_z}{\partial r} & 0 & 0
\end{bmatrix}, \ 
\bm{\mathsf{W}} =
\begin{bmatrix}
0 & \displaystyle -\frac{1}{r} \frac{\partial}{\partial r} \left( r U_\theta \right) & \displaystyle -\frac{\partial U_z}{\partial r} \\
\displaystyle \frac{1}{r} \frac{\partial}{\partial r} \left( r U_\theta \right) & 0 & 0 \\
\displaystyle \frac{\partial U_z}{\partial r} & 0 & 0
\end{bmatrix}.
\label{eq:44}
\end{align}
In this matrix form, the tensor $\bm{\mathsf{C}}$ defined by Eq.~(\ref{eq:22}) can be calculated as $\bm{\mathsf{C}} = \bm{\mathsf{S}} \bm{\mathsf{W}} + (\bm{\mathsf{S}} \bm{\mathsf{W}})^\mathrm{t} (= \bm{\mathsf{S}} \bm{\mathsf{W}} - \bm{\mathsf{W}} \bm{\mathsf{S}})$. Although the model expression of the Reynolds stress seems to be complicated, as seen in Eq.~(\ref{eq:23}), it can be easily calculated from the square root tensor as $\bm{\mathsf{R}} = \sqrt{\bm{\mathsf{R}}} \sqrt{\bm{\mathsf{R}}}^\mathrm{t}$ with Eq.~(\ref{eq:21}). This is another advantage of the usage of the square root tensor. We compare the results of the present model with the experimental data provided by Imao \textit{et al}. \cite{imaoetal1996}, where the bulk Reynolds number is set to $\mathrm{Re}_\mathrm{m} = 10000$. The rotation rate $N$ is defined as $N = U_{\theta,\text{wall}}/U_\mathrm{m}$ and experiments of three parameters, $N =0$, $0.5$, and $1$, were performed. 

\begin{figure}[htp]
\centering
 \begin{tabular}{c}
  \begin{minipage}{0.49\hsize}
   \centering
   \includegraphics[scale=0.65]{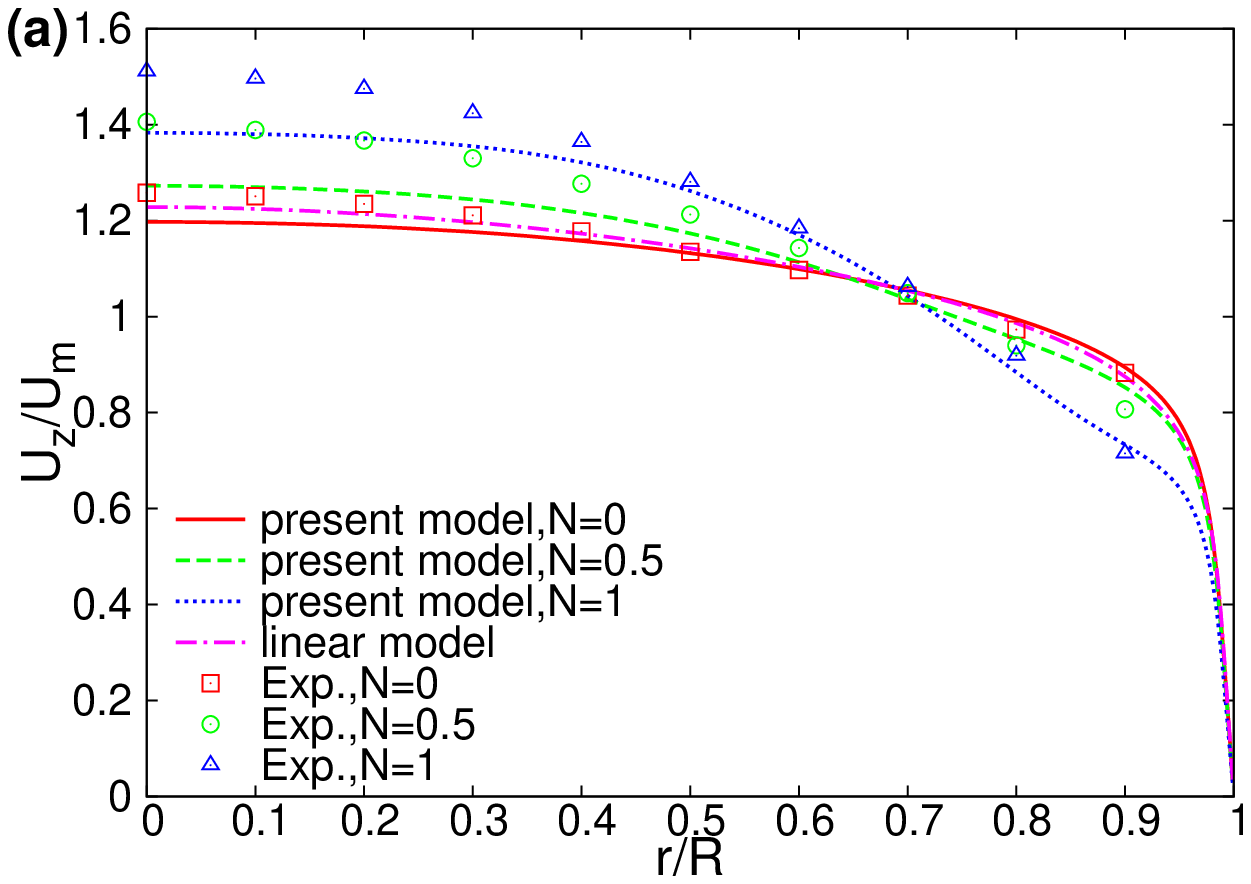}
  \end{minipage}

  \begin{minipage}{0.49\hsize}
   \centering
   \includegraphics[scale=0.65]{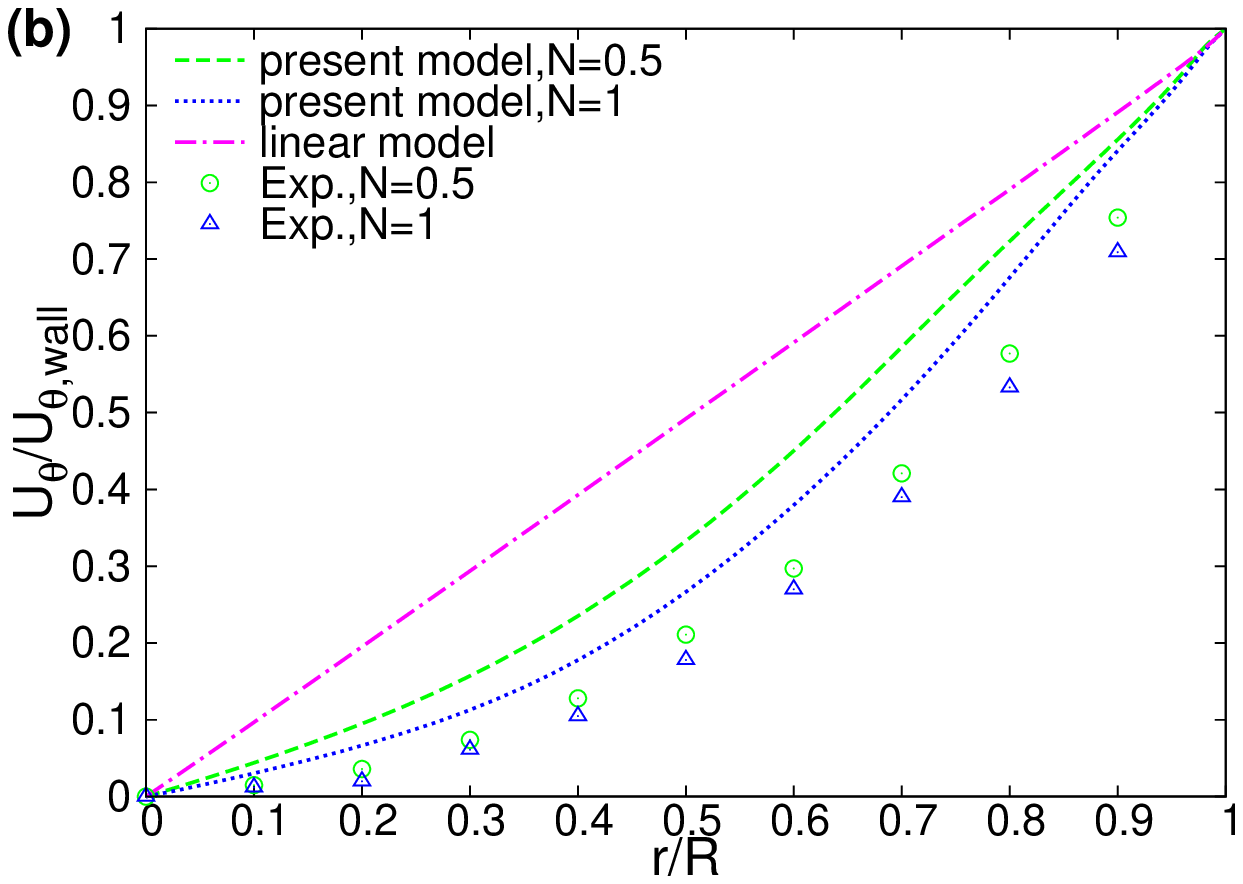}
  \end{minipage}
 \end{tabular}
\caption{Profiles of (a) the mean axial velocity $U_z/U_\mathrm{m}$ and (b) the mean swirl velocity $U_\theta / U_{\theta, \text{wall}}$ in the axially rotating turbulent pipe flow at $\mathrm{Re}_\mathrm{m} = 10000$.}
\label{fig:7}
\end{figure}

Figure~\ref{fig:7}(a) shows the profile of the mean axial velocity $U_z$ in a steady state for each rotation parameter. The result of the linear eddy-viscosity model is also plotted for reference. It is seen that the present model predicts the rotation-dependent property of the mean axial velocity; namely, the mean velocity gradient becomes steep as the rotation parameter increases. Figure \ref{fig:7}(b) shows the profile of the mean swirl velocity $U_\theta$ in a steady state for two rotating cases. Moreover, neither the linear eddy-viscosity model nor quadratic-nonlinear models can predict the mean swirl velocity \cite{syb2000}; they just predict the solid body rotation solution, $U_\theta = r U_{\theta,\text{wall}}/R$. As seen in Fig.~\ref{fig:7}(b), the present model predicts the rotation-dependent mean swirl velocity. Although both trends of the mean axial velocity and mean swirl velocity are reproduced by the present model, there is still room for improvement in the quantitative sense. This point should be further discussed in the future. However, it should be rather emphasized that the present model is a realizable quartic-nonlinear ARSM which has never been proposed. It can reproduce a feature of three-dimensional flow owing to the quartic nonlinearity on the mean velocity gradient with ensuring the realizability conditions.

\section{\label{sec:level5}Conclusions}

In this study, the realizable ARSM based on the square root tensor of the Reynolds stress \cite{ariki2015sqrt} was further developed. By imposing the realizability conditions, we could develop a physically consistent and numerically stable turbulence model. In contrast to the conventional method, the square root modeling provides a systematic and simpler construction of the realizable model, even when it involves higher-order nonlinearity on the mean velocity gradient. The cubic or higher-order nonlinearity on the mean velocity gradient is required to predict three-dimensional flows, which denote the turbulent flows with three-dimensional mean velocity. In fact, the cubic nonlinearity is required to predict the mean swirl flow in an axially rotating turbulent pipe flow \cite{syb2000}.

For the development from the previous study by Ariki \cite{ariki2015sqrt}, we proposed a systematic way to determine the dimensional coefficients of the model. Consequently, a dependence of the eddy viscosity on the mean strain and rotation rate is deductively introduced. Moreover, the dimensional coefficients depend on the other additional terms chosen for an expansion basis of the square root tensor. In this sense, the present modeling gives a generic way to determine a functional form of the dimensional coefficients. 

As a simple model applicable to three-dimensional flows, a quartic-nonlinear ARSM was proposed through the square root tensor. The performance of the proposed model was numerically verified in a turbulent channel flow, a homogeneous turbulent shear flow, and an axially rotating turbulent pipe flow. The present model gave reasonable predictions for both mean velocity profile in the turbulent channel flow and time evolution of the turbulent energy in the homogeneous turbulent shear flow. Moreover, the rotation rate-dependent property of the mean axial and swirl velocity in the axially rotating turbulent pipe flow was predicted, which cannot be predicted by any quadratic-nonlinear eddy-viscosity model, including the conventional realizable models \cite{szl1993,szl1995,akn1997,hamba2001,ajl2003}. 

The present model overestimated the anisotropy of the Reynolds stress in the homogeneous turbulent shear flow and underestimated the mean axial and swirl velocity in the axially rotating turbulent pipe flow; namely, there remains enough room for further improvement in the quantitative sense. For simplicity of the model, we did not consider the effect of the turbulent helicity \cite{yy1993,yb2016,inagakietal2017} and the time history effect of the mean strain rate \cite{speziale1987,hd2008,ariki2019} in this study. Hence, the model can be further developed. However, note that the present model is a quartic-nonlinear ARSM that always satisfies the realizability conditions, which has never been proposed. Such a higher-order realizable turbulence model, involving quartic nonlinearity on the mean velocity gradient, is expected to be useful in numerically stable predictions of turbulent flows with three-dimensional mean velocity.

\appendix

\makeatletter
\renewcommand{\theequation}{\thesection\arabic{equation}}
\@addtoreset{equation}{section}
\makeatother

\section{\label{sec:a}Physical interpretation of the square root tensor}

Let us introduce an isotropic stochastic vector $\xi_i$, which satisfies $\langle \xi_i \rangle = 0$ and $\langle \xi_i \xi_j \rangle = \delta_{ij}$. We consider the anisotropic velocity fluctuation $u_i'$ expressed by the following mapping form:
\begin{align}
u_i'  = \sqrt{R}_{ij} \xi_j,
\label{eq:a1}
\end{align}
where $\sqrt{R}_{ij}$ denotes the anisotropic mapping operator. Note that we consider a non-fluctuating $\sqrt{R}_{ij}$; namely, $\langle \sqrt{R}_{ij} \rangle = \sqrt{R}_{ij}$ and $\langle \sqrt{R}_{ij} \xi_j \rangle = \sqrt{R}_{ij} \langle \xi_j \rangle = 0$. Then, the Reynolds stress is calculated by its definition, $R_{ij} = \langle u_i' u_j' \rangle$, as
\begin{align}
R_{ij} = \left< \sqrt{R}_{ia} \xi_a \sqrt{R}_{jb} \xi_b \right> = \sqrt{R}_{ia} \sqrt{R}_{jb} \left< \xi_a \xi_b \right>
= \sqrt{R}_{ia} \sqrt{R}_{ja}.
\label{eq:a2}
\end{align}
This is just the definition of the square root tensor of the Reynolds stress given by Eq.~(\ref{eq:10}). Hence, $\sqrt{R}_{ij}$ can be interpreted as an envelope representing the anisotropy of the velocity fluctuation. 

Here, we discuss an expression for the velocity fluctuation in anisotropic turbulence in terms of the mean velocity gradient. For theoretical convenience, we expand the velocity fluctuation around the isotropic part as $u_i' = u^{(0)}_i + u^{(1)}_i$, where $u^{(0)}_i$ denotes the isotropic velocity fluctuation and $u^{(1)}_i$ denotes the anisotropic velocity fluctuation around $u^{(0)}_i$. The perturbation equation for $u^{(1)}_i$ from isotropic turbulence can be written as \cite{tsdia,yoshizawabook,yoshizawa1995}
\begin{align}
L_u u^{(1)}_i = - \frac{1}{2} ( S_{ij} + W_{ij} ) u^{(0)}_j,
\label{eq:a3}
\end{align}
where 
\begin{align}
L_u u^{(1)}_i & =
\frac{D u^{(1)}_i}{D t} + \frac{\partial}{\partial x_j} \left( u^{(1)}_i u^{(0)}_j + u^{(0)}_i u^{(1)}_j
- \left< u^{(1)}_i u^{(0)}_j \right> - \left< u^{(0)}_i u^{(1)}_j \right> \right) + \frac{\partial p^{(1)}}{\partial x_i} - \nu \nabla^2 u^{(1)}_i.
\label{eq:a4}
\end{align}
Equation~(\ref{eq:a3}) can be formally integrated as
\begin{align}
u^{(1)}_i = - \frac{1}{2} L_u^{-1} ( S_{ij} + W_{ij} ) u_j^{(0)}.
\label{eq:a5}
\end{align}
This inverse operation $L_u^{-1}$ denotes the time integration with a characteristic time scale of turbulence and we approximate this integration as \cite{yoshizawa1995,yoshizawabook}
\begin{align}
u^{(1)}_i = - \frac{1}{2} L_u^{-1} ( S_{ij} + W_{ij} ) u^{(0)}_j = - \frac{1}{2} \tau ( S_{ij} + W_{ij} ) u^{(0)}_j,
\label{eq:a6}
\end{align}
where $\tau$ represents a characteristic time scale of turbulence. When the isotropic velocity fluctuation is written as $u^{(0)}_i = \sqrt{2K_0/3} \xi_i$, the velocity fluctuation $u_i'$ reads
\begin{align}
u_i' = u^{(0)}_i + u^{(1)}_i 
= \sqrt{\frac{2K_0}{3}} \left[ \delta_{ij} - \frac{1}{2} \tau ( S_{ij} + W_{ij} ) \right] \xi_j.
\label{eq:a7}
\end{align}
Comparing Eqs.~(\ref{eq:a1}) and (\ref{eq:a7}), the square root tensor $\sqrt{R}_{ij}$ can be expressed as
\begin{align}
\sqrt{R}_{ij}
= \sqrt{\frac{2K_0}{3}} \left[ \delta_{ij} - \frac{1}{2} \tau ( S_{ij} + W_{ij} ) \right].
\label{eq:a8}
\end{align}
This expression of $\sqrt{R}_{ij}$ corresponds to Eq.~(\ref{eq:17}), where $\gamma_0 = \sqrt{2K_0/3}$, $\gamma_S = \gamma_0 f_S$, $\gamma_W = \gamma_0 f_W$, $\gamma_N = 0$, and $f_S = f_W = \tau/2$. The higher-order nonlinear term can be obtained by considering the higher-order perturbation. Consequently, the modeling based on the square root tensor is interpreted as the modeling based on the generation mechanism of the anisotropy of the velocity fluctuation.

\bibliography{ref}

\end{document}